\documentclass{appolb}
\usepackage{graphicx}

\begin{document}
\title{Neutrino Astrophysics.%
\thanks{Lectures presented at 52th Winter School of Theoretical Physics}%
}
\author{Cristina Volpe
\address{Astro-Particule et Cosmologie (APC), CNRS, Universit\'e Denis Diderot,\\ 10, rue Alice Domon et L\'eonie Duquet, 75205 Paris Cedex 13, France.}}
\maketitle
\begin{abstract}
We summarize the progress in neutrino astrophysics and emphasize  open issues in our understanding of neutrino flavor conversion in media.  We discuss solar neutrinos, core-collapse supernova neutrinos and conclude with ultra-high energy neutrinos.
\end{abstract}
  
\section{Introduction}
\noindent
Nature has provided us with a variety of neutrino sources, from the not yet observed  1.9 K cosmological background to the IceCube PeV neutrinos \cite{Aartsen:2014gkd}, whose origin is still mysterious. 
Neutrinos are intriguing weakly interacting particles. 
After 1998 many unknown properties have been determined thanks to the discovery of neutrino oscillations, first proposed in \cite{Pontecorvo:1957cp} and observed by the Super-Kamiokande experiment using atmospheric neutrinos \cite{Fukuda:1998mi}. This discovery is fundamental for  particle physics, for astrophysics and for cosmology.

Neutrino oscillations is an interference phenomenon among the $\nu$ mass eigenstates, that occurs if neutrinos are massive and if the mass (propagation basis) and the flavor (interaction basis) do not coincide. The Maki-Nakagawa-Sakata-Pontecorvo matrix relates these two basis \cite{Maki:1962mu}.
Within three active flavors, such a matrix depends on three mixing angles, one Dirac  and two Majorana CP violating phases.
In the last two decades solar, reactor and accelerator experiments have precisely determined most of the oscillation parameters, including the so-called  atmospheric 
$\Delta m_{23}^2 = m_{3}^2 -  m_{2}^2 = 7.6 \times 10^{-3} $eV$^2$,  and solar $\Delta m_{12}^2 = m_{2}^2 -  m_{1}^2 = 2.4 \times 10^{-5} $eV$^2$ mass-squared differences \cite{Agashe:2014kda}.
Moreover the sign of $\Delta m^2_{12}$ has been measured since $^{8}$B neutrinos undergo
the Mikheev-Smirnov-Wolfenstein (MSW) effect   \cite{Wolfenstein:1977ue,Mikheev:1986gs} in the Sun \cite{Ahmad:2002jz,Eguchi:2002dm,Robertson:2012ib}.
The sign of $\Delta m_{23}^2$ is still unknown, either $\Delta m_{31}^2 > 0$  and
the lightest mass eigenstate is $m_1$ (normal ordering or "hierarchy"), or  $\Delta m_{31}^2 < 0$  it is $m_3$ (inverted ordering).
Most of neutrino oscillation experiments can be interpreted within the framework of three active neutrinos. However a few measurements present anomalies that require further clarification. 
Sterile neutrinos that do not couple to the gauge bosons but mix with the other active
species could be the origin of the anomalies. Upcoming experiments such as STEREO or CeSox will cover most of the mixing parameters identified in particular by the "reactor anomaly"
\cite{Mention:2011rk}. 

Among the fundamental properties yet to be determined are the mechanism for the neutrino mass, the absolute mass value and ordering, the neutrino nature (Dirac versus Majorana), the existence of CP violation in the lepton sector and of sterile neutrinos.
The combined analysis of available experimental results shows a preference for normal ordering and for a non-zero CP violating phase, currently favouring $\delta = 3 \pi/2$, although statistical significance is still low \cite{Marrone:Nu2016}. 
 In the coming decade(s) experiments will aim at determining the mass ordering, the Dirac CP violating phase, the neutrino absolute mass and hopefully nature as well.
 Moreover Super-Kamiokande with Gadolinium should have the sensitivity to discover the relic supernova neutrino background \cite{Xu:2016cfv}. 
 
\section{Solar neutrinos}
\noindent
Electron neutrinos are constantly produced in our Sun and in low mass main sequence stars through the proton-proton (pp) nuclear reaction chain that produces 99 $\%$ of their energy by burning hydrogen into helium-4  \cite{Bethe:1939bt}. The corresponding solar neutrino flux receives contributions from both fusion reactions and beta-decays of $^{7}$Be and $^{8}$B (Figure 1). First measured by R. Davis pioneering experiment \cite{Davis:1968cp}, such flux was found to be nearly a factor of three below predictions \cite{Bahcall:1968hc}. Over the decades solar neutrino experiments  have precisely measured electron neutrinos from the different pp branches, usually referred to as the pp, pep, $^{7}$Be and $^{8}$B and hep neutrinos. 
The measurement of a reduced solar neutrino flux, compared to standard solar model predictions (the so-called the "solar neutrino deficit problem"), 
has been confirmed by experiments mainly sensitive to electron neutrinos, but with some sensitivity to the other flavors. 

The advocated solutions included unknown neutrino properties (e.g. flavor oscillations, a neutrino magnetic moment coupling to the solar magnetic fields, neutrino decay, the MSW effect) and questioned the standard solar model. In particular, the MSW effect is due to the neutrino interaction with matter while they traverse a medium. 
\begin{figure}
\begin{center}
\includegraphics[width=0.7\textwidth]{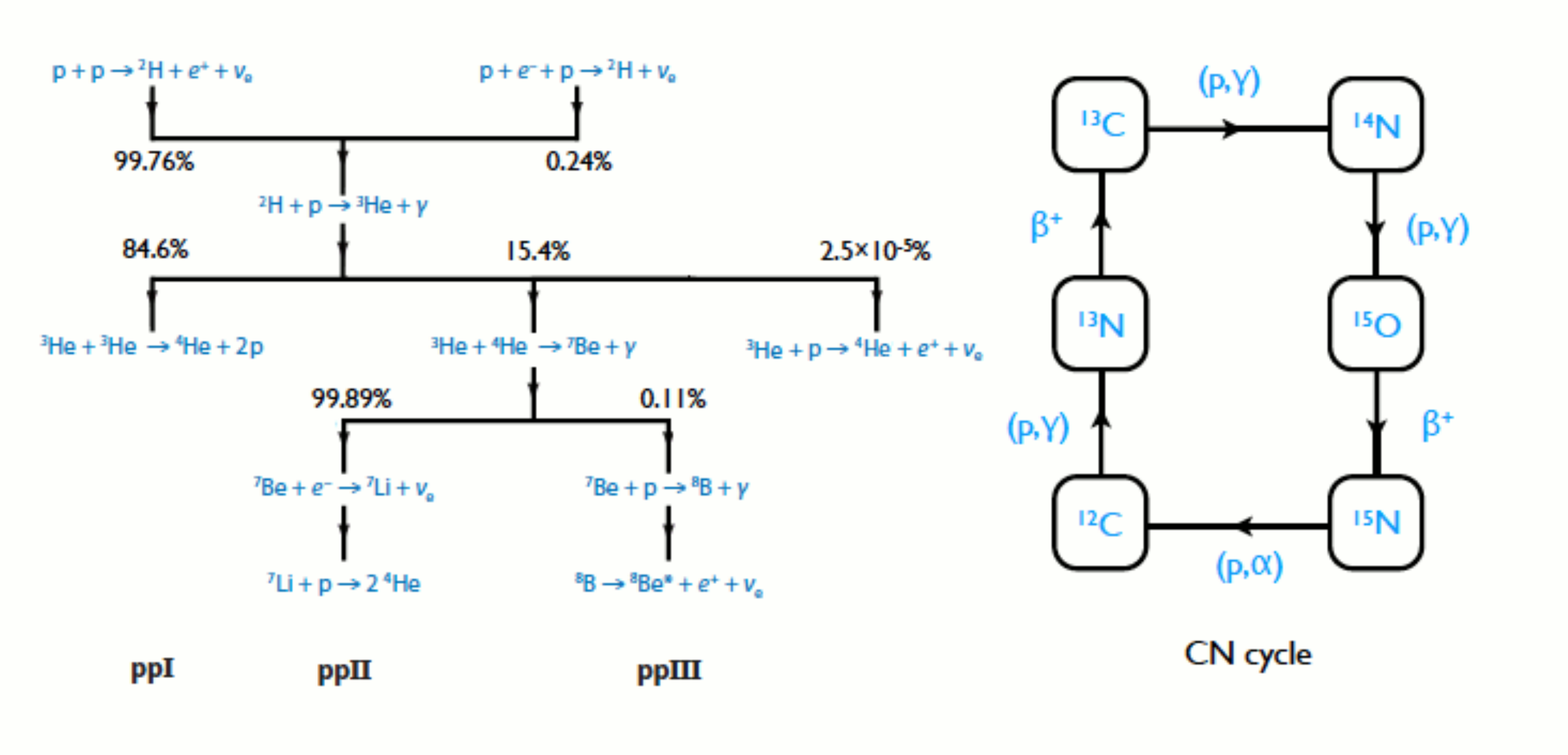}
\caption{The left figure shows  the proton-proton (pp) nuclear reaction chain with its three branches. The pp chain is responsible for energy production in our Sun and low mass stars. The theoretical branching percentages define the relative rates of the competing reactions. The right figure shows the CN I cycle which is thought to play an important role for energy production in massive stars. The $^{15}$O and $^{13}$N neutrinos have not been observed yet \cite{Robertson:2012ib}.}
\label{fig1}
\end{center}
\end{figure}

The solar puzzle is definitely solved by the discovery of the neutrino oscillation phenomenon  \cite{Fukuda:1998mi} and the results obtained by the SNO and KamLAND experiments (see \cite{Robertson:2012ib} for a review on solar neutrino physics). In fact, using elastic scattering, charged- and neutral- current neutrino interactions on heavy water, the SNO experiment has showed that the measurement of the total $^{8}$B solar neutrino flux is consistent with the predictions of the standard solar model : solar electron neutrinos convert into the other active flavors. In particular, the muon and tau neutrino components of the solar flux has been measured at 5 $\sigma$ \cite{Ahmad:2002jz}. Moreover the reactor experiment KamLAND has definitely identified the Large Mixing Angle (LMA) solution, by observing reactor electron anti-neutrino disappearance at an average distance of 200 km \cite{Eguchi:2002dm}. The ensemble of these observations shows that low energy solar neutrinos are suppressed by averaged vacuum oscillations while neutrinos having more than 2 MeV energy 
are suppressed because of the MSW effect (Figure 2). Theoretically one expects $P (\nu_e \rightarrow \nu_e)  \approx 1 - {1 \over 2} \sin^2 2 \theta_{12} \approx 0.57 $
(with $\theta_{12} = 34^{\circ}$) for ($< 2$ MeV) solar neutrinos; 
for high energy portion of the $^{8}$B spectrum, the matter-dominated survival probability is $P (\nu_e \rightarrow \nu_e) ^{high~density}  \rightarrow \sin^2\theta_{12} \approx 0.31$ (see \cite{Robertson:2012ib}). The precise determination of the transition between the vacuum averaged and the LMA solution brings valuable information since deviations from the simplest vacuum-LMA transition could point to new physics, such as non-standard neutrino interactions \cite{Friedland:2004pp}.   

\begin{figure}
\begin{center}
\includegraphics[width=0.7\textwidth]{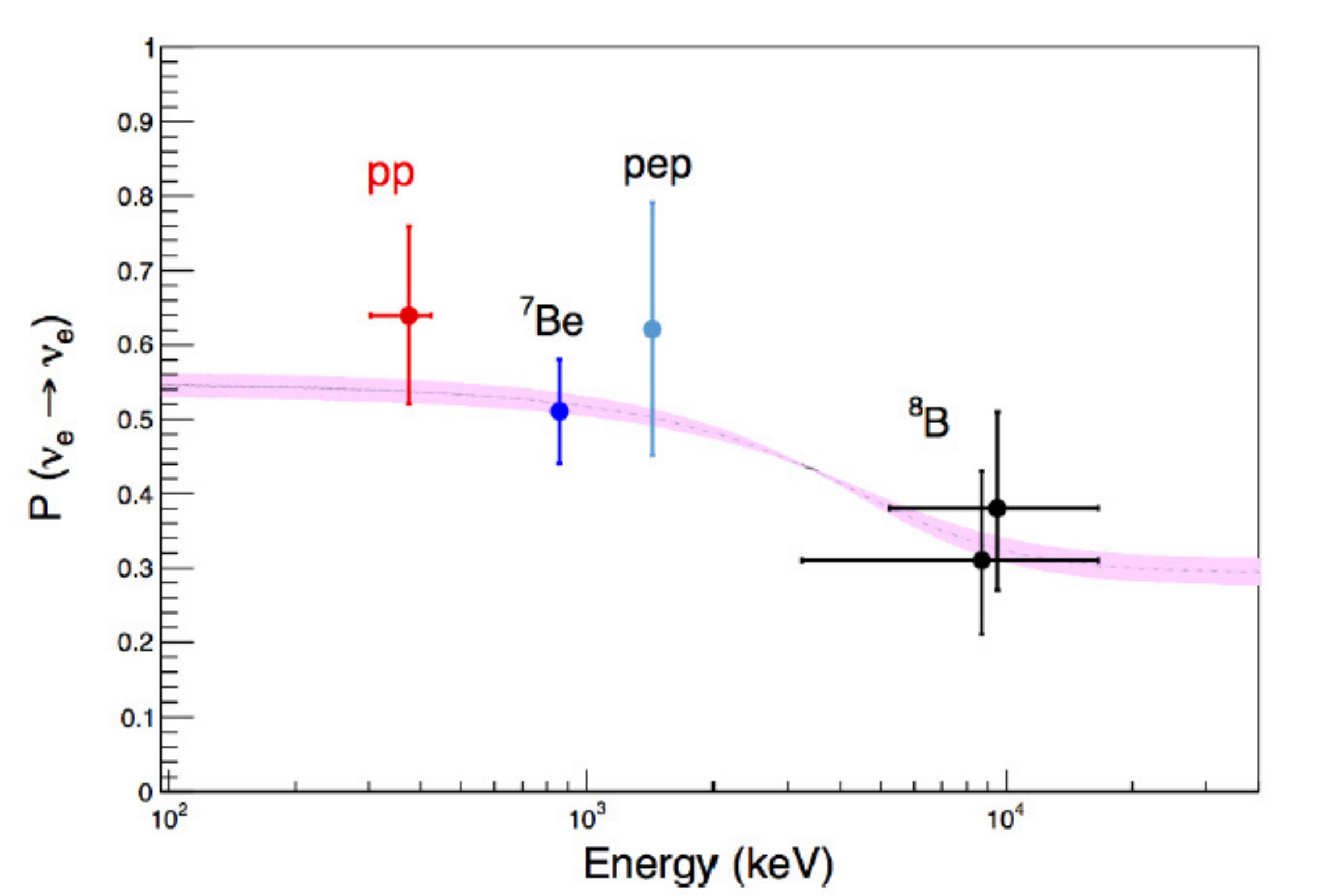}
\caption{Electron neutrino survival probability, as a function of the neutrino energy, for the pp, pep, $^{7}$Be, $^{8}$B neutrinos from the Borexino experiment. The results are compared to averaged vacuum oscillation prediction ($E_{\nu} < 2$ MeV) and the MSW prediction ($E_{\nu} > 2$ MeV), taking into account present uncertainties on mixing angles. Figure from \cite{Bellini:2014uqa}. }
\label{fig1}
\end{center}
\end{figure}
The Borexino experiment has precisely measured the low energy part of the solar neutrino flux, namely the pep \cite{Collaboration:2011nga}, $^{7}$Be \cite{Arpesella:2008mt}. Moreover, by achieving challenging reduced backgrounds, the collaboration has reported the first direct measurement
of pp neutrino, the keystone of the fusion process in the Sun. The measured flux is consistent with the standard solar model predictions  \cite{Bellini:2014uqa}. 

The ensemble of solar observations has established that the Sun produces $3.84 \times 10^{33}$ ergs/s via the pp chain. Moreover the occurrence of the MSW effect for the high energy solar neutrinos shows that these particles change flavor in vacuum in a very different way than in matter. In fact, in the central high density regions of the star 
the flavor  coincide with the matter eigenstates. During their propagation towards the edge of the Sun, they encounter a resonance (if the MSW resonance condition is fulfilled) and evolve adiabatically through it depending on the neutrino energies, squared-mass difference value and the gradient of the matter density. 
Adiabaticity  implies that the matter eigenstates mixing is suppressed at the resonance.
In the latter case  electron neutrinos can efficiently convert into muon and tau neutrinos. 
The MSW phenomenon is analogous to the two-level system in quantum mechanics. 
It occurs in numerous contexts, including the early universe (at the epoch of the primordial elements formation), massive stars like core-collapse supernovae, accretion disks aroung black holes and the Earth. 

Future measurements will aim at observing solar neutrinos produced in the Carbon-Nitrogen-Oxygen (CNO) cycle which is thought to be the main mechanism for energy production in massive main sequence stars  \cite{Bethe:1939bt}. Borexino experiment has provided the strongest constraint on the CNO cycle which represents 1 $\%$ of energy production in the Sun, consistent with standard solar model predictions \cite{Collaboration:2011nga}. The achievement of increased purity both by Borexino and  SNO+ could allow to reach the sensivity for this challenging measurement. 

Beyond furnishing confirmation of stellar evolutionary models,
the observation of CNO neutrinos could help solving the so-called solar "opacity" problem. Standard solar models predict solar neutrino fluxes, from the pp cycle, in agreement with observations. However the GS98-SFII and AGSS09-SFII models differ for their treatment of the metal element contributions (elements heavier than He).
The first model uses older abundances for volatile elements that are obtained by an absorption line analysis in which the photosphere is treated as one-dimensional that yields a metallicity of $(Z/X)_S = 0.0229$ ($Z$ and $X$ being the metal and hydrogen abundances respectively) with solar fusion II cross sections (GS98-SFII). The second model takes abundances from a three-dimensional photospheric model 
$(Z/X)_S = 0.0178$ (AGSS09-SFII). The latter produces a cooler core by $1 \%$ and lower fluxes of temperature sensitive neutrinos such as $^{8}$B ones. A comparison of the solar parameters used in  the two models and corresponding predictions on the neutrino fluxes are given in Tables 1 and 2 of Ref.\cite{Robertson:2012ib}.
The "solar opacity problem" is the inconsistency between AGSS09 that uses the best description of the solar photosphere, while GS98 has the best agreement with helioseimic data that are sensitive to the interior composition.  Since there is approximately 30 $\%$ difference between C and N abundances in the two models,
a measurement of CNO neutrinos with $12 \%$ precision, which could be achieved in the future, will allow to determine the solar opacity.

\section{Supernova neutrinos}
\subsection{Core-collapse supernovae and SN1987A}
\noindent
Core-collapse Supernovae (SNe) are stars with mass $M > 6~M_{sun}$ ($M_{sun}$ being the Sun's mass) whose core undergoes gravitational collapse at the end of their life.  These include types II and Ib/c depending on their spectral properties.
They are of type II if they exhibit H lines in their spectra and of type I if they don't because the star has lost the H envelope. SNe IIb have a thin H envelope; type II-P and II-L present a plateau or a linear decay of the light curves after the peak.    
The SNe Ib shows He and Si lines, while SNe Ic shows none of these indicating that before collapse the star has lost both the H envelope and He shells.  
The supernova can still appear as bright if the H envelope is present, otherwise it can be invisible (Type Ib/c)  \cite{Heger:2002by}.

In 1960 Hoyle and Fowler proposed that stellar death of SNII and I/b happens because of the implosion of the core \cite{Hoyle:1960zz}. The same year Colgate and Johnson suggested that a bounce of the neutron star forming launches a shock that ejects the matter to make it unbound \cite{Colgate:1960zz} (the "prompt model").
It was realised by Colgate and White \cite{Colgate:1966ax} that a gravitational binding energy of the order of $E \approx G M_{NS}^2/R_{NS} > 10^{53}$ erg associated with the collapse of the star core to a neutron star (NS) would be released as neutrinos that would deposit energy to trigger the explosion. 
Arnett \cite{Arnett1966} and Wilson \cite{Wilson1971} critized the model because it would not give enough energy. 
Wilson revisited the model and developed it further :  the ejection of the mantle would be preceded by an accreting phase in the so-called 
"delayed neutrino-heating mechanism" \cite{Bethe:1984ux}.

The fate of a massive star is mainly determined by the initial mass and composition and the history of its mass loss. Their explosion produces either neutron stars or black holes directly or by fallback.
Their initial masses range from 9 to 300 solar masses ($M_{sun}$). Stars having 6-8 $M_{sun}$ develop an O-Ne-Mg core while those with $M > 8$ $M_{sun}$  possess an iron core before collapse.  Hypernovae are asymmetric stellar explosions with high ejecta velocities, they are very bright, producing a large amount of Nickel. They are often associated with long-duration gamma-ray-bursts. Collapsars are all massive stars whose core collapse to a black hole and that have sufficient angular momentum to form a disk (see e.g. \cite{Heger:2002by,Janka:2012wk}). 

\begin{figure}\label{fig:2}
\begin{center}
\includegraphics[width=0.6\textwidth]{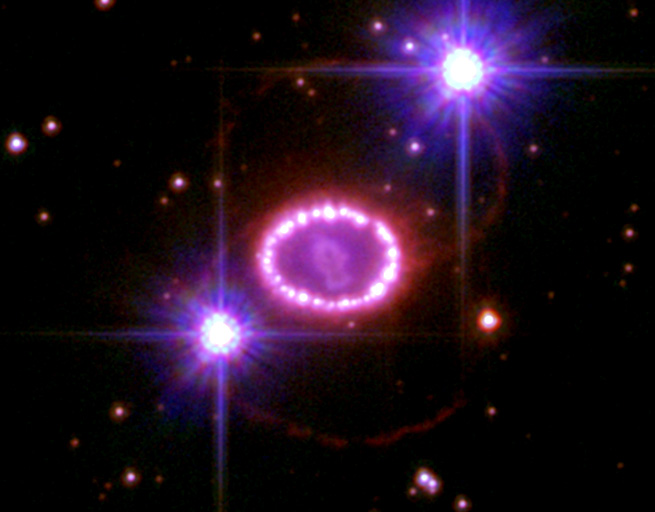}
\caption{Picture of SN1987A 20 years after its explosion with its three rings. The inner blowing ring was formed 20000 years before the explosion \cite{HTS}.}
\label{fig:sn1987}
\end{center}
\end{figure}

On 23 February 1987 Sk -$69^{\circ} 202$ exploded producing SN1987A, the first naked-eye supernova since Kepler's one in 1604. It was located in the Large Magellanic Cloud, a satellite galaxy of the Milky Way. The determined distance is 50 kpc from the Earth based on the expanding photosphere method from different groups which agree within 10 $\%$ (see Table I of \cite{Schmidt:1992yr}). This method to establish extragalactic distances allows to cover a wide range, from 50 kpc to 200 Mpc. 
From the observed light-curve and simulations it appears that the core mass of SN1987A progenitor was around 6 $M_{sun}$ and total mass $\approx$ 18 $M_{sun}$ and the progenitor radius  about $10^{12}$ cm \cite{Pod:92}. 
SN1987A is unique because it was observed in all wavelengths from gamma rays to radio, and for the first time, neutrinos were observed from the collapse of the stellar core. The neutrinos was first discovered  by Kamiokande II \cite{Hirata:1987hu}, then by IMB \cite{Bionta:1987qt} and Baksan \cite{Alekseev:1988gp}. The number of detected electron anti-neutrinos events were 16 in Kamiokande II, 8 in IMB and 5 in Baksan. Time, energy, SN-angle  and background rate for all the events is given in Table I of the recent review \cite{Vissani:2014doa}. Several hours before 5 events were seen in LSD detector that could be due to a speculative emission phase preceding the ones seen in the other detectors  \cite{Aglietta:1987it}. Such events are often discarded in the analysis of SN1987A data since their are object of debate. 
The earliest observations of optical brightening were recorded 3 hours  after neutrino's arrival.  An enthusiastic description of SN1987A discovery is reported in \cite{Suzuki:2008zzf}. 

Three puzzling features concerning SN1987A has set constraints on stellar evolutionary models and supernova simulations. The progenitor was a blue supergiant rather than a red supergiant while type II supernovae were thought to be produced by red supergiants. Large-mixing processes had transported radioactive nuclei from the deep core far into the H envelope of the progenitor and in the pre supernova ejecta, producing anomalous chemical abundances. The presence of  three ring-like geometry of the circumstellar nebula around the supernova (Figure \ref{fig:sn1987}) was implying a highly non-spherical structure of the progenitor envelope and its winds \cite{Pod:92}. Various explanations have been suggested for the presence of these rings, the inner one being dated  20000 years before the explosion. They might  have originated by a binary merger event of that epoch \cite{Pod:92, Podsiadlowski:2007zz} showing that rotation might have played a significant role in the dying star. However the prolate deformation of the supernova ejecta at the center of the ring system might have a very different origin (Figure \ref{fig:sn1987}).
In fact, the presence of large-mixing and the asymmetric ejecta indicates breaking of spherical symmetry due to hydrodynamical instabilities such as the bipolar Standing Accretion Shock Instability (SASI)  \cite{Janka:2007yu}.  SN1987A remnant is likely not a black hole since the progenitor was light enough to be stabilised by nuclear equation-of-states consistent with measured neutron star masses \cite{Janka:2007yu,Sato:1987yi}. There is currently no sign as well of a bright pulsar as the one  born from the supernova explosion  in the Crab nebula in 1054.

SN1987A neutrino observations have been used to derive constraints on fundamental physics and the properties of neutrinos, axions, majorons, light supersymmetric particles
and on unparticles. These are derived by the absence of non-standard signatures, by using the intrinsic neutrino signal dispersion or by the cooling time of the newborn neutron star.  Many such limits have been superseded by direct measurements with controlled sources on Earth, while other remain valuable constraints. For example, from the three hours delay in the transit time of  neutrinos and photons a tight limit can be on the difference between the speed of neutrinos $c_{\nu}$ and light $c$ is obtained, i.e.  $\mid (c_{\nu} - c)/c \mid  < 2 \times 10^{-9}$  \cite{Longo:1987ub}. 

SN1987A neutrinos have also confirmed the basic features of core-collapse supernova predictions concerning the neutrino fluence (time-integrated flux) and spectra.  From a comparative analysis of the observed neutrino events one gets as a best fit point $E = 5 \times 10^{52}$ erg and $T = 4$ MeV for the total gravitational energy radiated in electron anti-neutrinos and their temperature respectively
\cite{Vissani:2014doa}. Accoding to expectations, 99 $\%$ of the supernova gravitational binding energy should be converted in $\nu_e, \nu_{\mu}, \nu_{\tau}$ neutrinos (and anti-neutrinos) in the several tens of MeV energy range. Such neutrinos are produced by pair annihilation, electron capture and neutron bremstrahlung
-- $e^- + (A,Z) \rightarrow \nu_e + (A, Z-1)$, $\nu + N \rightarrow \nu + N$, $(A,Z) + \nu \rightarrow (A,Z) + \nu $, $e^+ + e^- \rightarrow \nu + \bar{\nu}$, $N + N \rightarrow N + N + \nu + \bar{\nu}$. If one considers that energy equipartition among the neutrino flavors is rather well satisfied, one gets about $3 \times 10^{53}$ ergs and the emission time is also found to be of $15$ s. Considering that the neutrino spectra are to a fairly good approximation thermal, one gets for the average electron anti-neutrino energy $E_{\nu} = 3 T $ giving 12 MeV at the best fit point. This appears currently more compatible with supernova simulations based on realistic neutrino transport, although it has appeared much lower than the expected value of $15$ MeV claimed for a long time.  

Supernova neutrinos are tightly connected with two major questions in astrophysics, namely what is the mechanism that makes massive stars explode and what is (are) the site(s) where the heavy elements are formed through the so-called rapid neutron capture process (or $r$-process). Neutrinos would contribute in neutrino-driven winds in core-collapse supernovae, accretion-disks around black holes and neutron-star mergers. In fact, the interaction of electron neutrinos and anti-neutrinos with neutrons and protons in such environments determines the neutron-to-proton ratio, a key parameter of the r-process. Obviously astrophysical conditions and the properties of exotic nuclei (like masses, $\beta$-decay half-lives or fission) are crucial in determining the abundances.  
Several studies  have shown that neutrinos impact the neutron richness of a given astrophysical environment. Finally assessing their influence still requires extensive simulations (see the reviews in the Focus Issue \cite{Volpe:2014yqa}). 

Various mechanisms for the SN blast are investigated, including a thermonuclear, a bounce-shock, a neutrino-heating,  a magnetohydrodynamic, an acoustic and a phase-transition mechanisms (see  \cite{Janka:2012wk}). Since the kinetic energy in SN events goes from $10^{50-51}$ erg for SNe up to several $10^{52}$ erg for hypernovae, the explosion driving mechanism have to comply, among others, with providing such energies. The neutrino-heating mechanism with non-radial hydrodynamical instabilities (convective overturn with SASI) appear to be a good candidate to drive iron-core collapse supernova explosions; while the more energetic hypernovae events could be driven by the magnetohydrodynamical mechanism. Note that a new neutrino-hydrodynamical instability  termed LESA  (Lepton-number Emission Self-sustained Asymmetry) has been identified  \cite{Tamborra:2014aua}. Simulations of the lighter O-Ne-Mg core-collapse supernovae do explode, while this is not yet the case for iron-core collapse ones.
Successful explosions  for two-dimensional simulations with realistic neutrino transport have been obtained for several progenitors; while the first  three-dimensional explosion are being obtained \cite{Janka:2016fox}.
  
 \subsection{Neutrino flavor conversion in astrophysical environments} 
 \noindent
 Neutrino propagation in cosmological or astrophysical environments is often described using effective (iso)spins, neutrino amplitudes, the density matrix approach,
 the path-integral formalism or many-body Green's functions (see \cite{Giunti:2007ry} and \cite{Volpe:2015rla} for a review). Note that the spin formalism gives a geometrical representation of neutrino evolution in flavor space. Here we briefly describe how to derive neutrino evolution equations useful for astrophysical applications based on the mean-field approximation. To this aim we use the density matrix formalism and follow the derivation in Ref.\cite{Serreau:2014cfa}.
  
\subsubsection{The evolution equations} 
\noindent
In the mass basis, at each time the spatial Fourier decomposition of a Dirac neutrino field reads
\begin{equation} \label{e:field}
\psi_{i} (t,\vec{x}) =  \int_{\vec p,s} e^{i \vec p \cdot \vec{x}} \,\psi_{i} (t,\vec p,s),
\end{equation} 
with
\begin{equation} \label{e:field2}
\psi_{i} (t,\vec p,s)= a_{i}(t,\vec p,s)u_{i}({\vec p,s}) + b_{i} ^{\dagger}(t,-\vec p,s) v_{i} (-{\vec p,s}),
\end{equation} 
where we note $\int_{\vec p}\equiv \int {d^3 {p} \over{(2 \pi)^3}}$ and $\int_{\vec p,s}\equiv \int_{\vec p}\sum_s $. The Dirac spinors corresponding to mass eigenstates $i$ are normalized as (no sum over $i$)
\begin{equation}
 u^\dagger_i (\vec p,s)u_i(\vec p , s')=v^\dagger_i (\vec p,s)v_i(\vec p , s')=\delta_{ss'}.
\end{equation} 
The standard particle and antiparticle annihilation operators (in the Heisenberg picture) for neutrinos of mass $m_i$, momentum $\vec p$ and helicity $s$ satisfy the canonical equal-time anticommutation relations.
\begin{equation}\label{e:commutators1}
\{ a_{i}(t,\vec p,s), a^{\dagger}_{j}(t,\vec p\,',s') \} 
=  (2 \pi)^3 \delta^{(3)}(\vec p - \vec p\,')\delta_{ss'}\delta_{i j} 
\end{equation}
\begin{equation} 
 \{ a_{i}(t,\vec p,s), a_{j}(t,\vec p\,',s') \} 
=   \{ a^{\dagger}_{i}(t,\vec p,s), a^{\dagger}_{j}(t,\vec p\,',s') \} =0
\end{equation}
and similarly for the anti-particle operators. 

In the flavor basis, the field operator is obtained as 
\begin{equation}
 \psi_\alpha(t,\vec x)=U_{\alpha i}\,\psi_i(t,\vec x),
\end{equation}
with $U$ the Maki-Nakagawa-Sakata-Pontecorvo unitary matrix  \cite{Maki:1962mu}. Note that the indices can refer to active as well as to sterile neutrinos.
In the framework of three active neutrinos the three mixing angles of $U$ are now determined. Two are almost maximal, while the third one is small \cite{Agashe:2014kda}. 
The Dirac and Majorana CP violating phases are still unknown \cite{Agashe:2014kda}. 

The flavor evolution of a neutrino, or of an antineutrino, in a background can be determined using one-body density matrices, namely expectation values of bilinear products of creation and annihilation operators
\begin{equation} \label{e:rho}
\rho_{ij}(t,\vec q,h,\vec q\,'\!,h')  = \langle a^{\dagger}_{j}(t,\vec q\,'\!,h') a_{i}  (t,\vec q,h)  \rangle,
\end{equation} 
\begin{equation} 
\label{e:arho}
\bar{\rho}_{ij}(t,\vec q,h,\vec q\,'\!,h') = \langle b^{\dagger}_{i}(t,\vec q,h)   b_{j} (t,\vec q\,'\!,h')  \rangle,
\end{equation} 
where the brackets denote quantum and statistical average over the medium through which neutrino are propagating. 
For particles without mixings, only diagonal elements are necessary and relations (\ref{e:rho}-\ref{e:arho}) 
correspond to the expectation values of the number operators. If particles have mixings as is the case for neutrinos, 
the off-diagonal contributions ($i \neq j$) of $\rho$ and $\bar{\rho}$
account for the coherence among the mass eigenstates. 

The mean-field equations employed so far to investigate flavor evolution in astrophysical environments evolve the particle and anti-particle correlators $\rho$ and $\bar{\rho}$.
However, the most general mean-field description 
includes further correlators. First,  densities with "wrong" helicity states, such as 
\begin{equation} \label{e:pm}
\rho_{ij}(t,\vec q,-,\vec q\,'\!,-)  = \langle b^{\dagger}_{j}(t,\vec q,-)   a_{i} (t,\vec q\,'\!,-) \rangle
\end{equation} 
are present. These have already been shown to impact neutrino evolution in presence of 
magnetic fields \cite{Lobanov:2001ar,deGouvea:2012hg}. They also give non-zero contributions if non-zero 
mass corrections are included \cite{Vlasenko:2013fja,Vlasenko:2014bva}.
Moreover two-point correlators called abnormal or pairing densities \cite{Serreau:2014cfa,Volpe:2013uxl}
\begin{equation} 
\label{e:kappa}
\kappa_{ij}(t,\vec q,h,\vec q\,'\!,h') = \langle b_{j}(t,\vec q\,'\!,h')   a_{i} (t,\vec q,h)  \rangle, 
\end{equation} 
and the hermitian conjugate also exist. Equations of motion including them have first been derived in Ref.\cite{Volpe:2013uxl}. 
If neutrinos are Majorana particles, correlators similar to (\ref{e:kappa}) can be defined, as done in Ref.  \cite{Serreau:2014cfa}, such as $\langle a_{j}(t,\vec q\,'\!,-)   a_{i} (t,\vec q,-)  \rangle$ or $\langle b^{\dagger}_{j} (t,\vec q\,'\!,+) b^{\dagger}_{i}(t,\vec q,+) \rangle$
that violate total lepton number. 
The most general mean-field evolution equations for Dirac or Majorana neutrinos evolving in an inhomogeneous medium is derived in Ref.\cite{Serreau:2014cfa}. 

The effective most general mean-field Hamiltonian takes the general bilinear form  ($\hbar=c=1$)
\begin{equation} \label{e:Heff}
H_{\rm eff}(t) = \int d^3 {x}\, \bar{\psi}_{i}(t,\vec{x})\Gamma_{ij}(t,\vec{x}) \psi_{j}(t,\vec{x}),
\end{equation} 
where $\psi_i$ denotes the $i$-th component of the neutrino field in the mass basis Eq.(\ref{e:field}). The explicit expression of the kernel $\Gamma$ depends on the kind of interaction considered (charged- or neutral-current interactions, non-standard interactions, effective coupling to magnetic fields, etc...). It does not need to be specified to obtain the general structure of the equations, but for practical applications. 

Equations of motion for the neutrino density matrix Eqs.(\ref{e:rho}-\ref{e:arho}) can be obtained from the Ehrenfest theorem:
\begin{equation} \label{e:Ehrenrho}
i \dot{\rho}_{ij}(t,\vec q,h,\vec q\,'\!,h') = \langle  [a^{\dagger}_{j}(t,\vec q\,',h')a_{i}(t,\vec q,h), H_{\rm eff}(t) ] \rangle 
\end{equation} 
and similarly for the other correlators. Spinor products can be introduced 
\begin{equation} 
\label{e:g1}
\Gamma_{ij}^{\nu\nu}(t,\vec q,h,{\vec q\,'\!,h'}) = \bar{u}_{i}(\vec q,h)\tilde\Gamma_{ij}(t,\vec q-\vec q\,')u_{j}({\vec q\,'\!,h'}),  
\end{equation} 
\begin{equation} \label{e:g2}
\Gamma_{ij}^{\bar\nu\bar\nu}(t,\vec q,h,\vec q\,'\!,h') = \bar{v}_{i}(\vec q,h)\tilde\Gamma_{ij}(t,-\vec q+\vec q\,')v_{j}({\vec q\,'\!,h'}), 
\end{equation} 
\begin{equation} \label{e:g3}
\Gamma_{ij}^{\nu\bar\nu}(t,\vec q,h,\vec q\,'\!,h') = \bar{u}_{i}(\vec q,h)\tilde\Gamma_{ij}(t,\vec q+\vec q\,')v_{j}({\vec q\,'\!,h'}) ,  
\end{equation} 
\begin{equation} \label{e:g4}
\Gamma_{ij}^{\bar\nu\nu}(t,\vec q,h,\vec q\,'\!,h') = \bar{v}_{i}(\vec q,h)\tilde\Gamma_{ij}(t,-\vec q-\vec q\,')u_{j}({\vec q\,'\!,h'}),
\end{equation} 
where the Fourier transform of the mean-field  in Eqs.(\ref{e:g1}-\ref{e:g4}) is defined as
\begin{equation} \label{e:fourier}
\Gamma_{ij}(t,\vec{x}) = \int_{\vec p}  e^{i \vec p \cdot \vec{x}}\, \tilde\Gamma_{ij}(t,\vec p\,).
\end{equation} 

If we neglect the contribution from pair-correlators, it is straightforward to show that the neutrino evolution for massive neutrinos propagating in a 
inhomogeneous medium is determined through the Liouville Von-Neumann equations of motion for the neutrino and antineutrino density matrices:
\begin{equation} \label{e:corr}
i \dot{\rho}(t)  =  \Gamma^{\nu\nu}(t)\cdot \rho(t) - \rho(t) \cdot \Gamma^{\nu\nu}(t) ,
\end{equation} 
\begin{equation} 
i \dot{\bar \rho}(t)  =  \Gamma^{\bar\nu\bar\nu}(t)\cdot\bar \rho(t) - \bar \rho(t) \cdot \Gamma^{\bar\nu\bar\nu}(t) ,
\end{equation}
Such equations become more explicit when one makes assumptions about the background. A common hypothesis, of interest for applications, is 
the one of a homogeneous, (an)isotropic, unpolarised medium. Such a condition correspond to
\begin{equation}\label{e:rhonuh}
\rho_{\vec{p}\,'h',\vec{p}h}= (2 \pi)^3 2 E_{p} \delta_{hh'} \delta^3 (\vec{p}  -\vec{p}\,') \rho_{\vec{p}},
\end{equation}
where $\rho$ here corresponds to the particle composing the background (electrons, neutrinos, etc \ldots).
Using Eqs.(\ref{e:Heff}-\ref{e:rhonuh}) one gets the following evolution equations for massless neutrinos :
\begin{equation} \label{e:corr}
i \dot{\rho}(t)  =  \left[h(t), \rho(t) \right]   ~~~~ i \dot{\bar \rho}(t)  =  \left[ \bar{h}(t), \bar{\rho}(t) \right] 
\end{equation} 
The mean-field equations for single-particle density matrices Eqs.(\ref{e:corr}) can be rigorously derived from the exact many-body description using the Born-Bogoliubov-Green-Kirkwood-Yvon (BBGKY) hierarchy by truncating the hierarchy to lowest order \cite{Volpe:2013uxl}.  
In absence of contributions from the neutrino mass, the pairing correlations and magnetic moments, 
the mean-field hamiltonian reduces to the well known form :
\begin{equation} 
\label{eq:scalar}
 h(t)=h^0+h^{\rm mat}(t)+h_{\nu\nu}(t)
\end{equation} 
where $h^0$ is the vacuum contribution, the second is the  neutrino-matter term  and the last comes from neutrino self-interactions\footnote{Note that in the case of our Sun, the neutrino self-interaction contribution is negligible and the medium is at a good approximation homogeneous and isotropic.}, whose contribution was first introduced by Pantaleone \cite{Pantaleone:1992eq}. 
The explicit expressions for the matter hamiltonian is 
\begin{equation} 
h^{\rm mat}(t) = \sqrt{2}G_F \left[N_e(t) - \frac{1}{2}N_n(t)\right],
\end{equation} 
with the particle number densities ($ f =e,n$ stands for electron and neutron) of the particles composing the medium
\begin{equation} 
 N_f(t)=2\int_{\vec p}\rho_f(t,\vec p\,)
\end{equation} 
and neutrino-neutrino interaction hamiltonian 
\begin{equation} 
h_{\nu\nu} =  \sqrt{2}G_F\!\!\int_{\vec p}\rho (t,{\vec q}) - \bar{\rho}(t,{-\vec q}).
\end{equation}  
The Hamiltonian for anti-neutrinos is the same as for neutrinos but with a different sign for the vacuum part, i.e. $ h(t)=- h^0+h^{\rm mat}(t)+h_{\nu\nu}(t)$.
Extended mean-field equations including contributions from wrong-helicity correlators such as 
Eq.(\ref{e:pm}) or from pair-correlators Eq.(\ref{e:kappa}) 
can be  cast in a compact matrix form  \cite{Volpe:2013uxl,Serreau:2014cfa} :
\begin{equation}\label{e:matrixform}
i\, \dot{\!{\cal R}} (t)= \left[ {\cal H}(t),{\cal R}(t)\right],
\end{equation} 
where ${\cal H}$ and ${\cal R}$ are the generalised Hamiltonian and density matrix respectively.
If neutrinos are Majorana particles, in these extended mean-field equations the neutrino and anti-neutrino sectors are coupled if mass, magnetic moments or 
pair correlators contributions are implemented. If neutrinos are Dirac particles, the wrong-helicity contributions from the mass (or in presence of magnetic moments) couple the neutrino (or anti-neutrino) sectors 
with the sterile one. With mass contributions only, ${\cal H}$ is a $N_f\times N_f$ scalar with flavour and helicity structure. Also the
neutrino-antineutrino mixing associated with the off-diagonal vector term of ${\cal H}$ gives a contribution perpendicular to the neutrino momentum. 
Therefore an anisotropic medium  is required for such contributions to be non-zero.

\subsubsection{Neutrino flavor conversion phenomena and open issues} 
\noindent
Important progress has been achieved in our understanding of how neutrinos change their flavor in massive stars, a case which is much more complex than the one of our Sun. The MSW effect in supernovae is well established. Since the star is very dense the MSW resonance condition can be fulfilled three times for
typical supernova density profiles \cite{Dighe:1999bi}. At high density the $\mu \tau$ resonance  depending on $(\theta_{23}, \delta m^2_{23})$ takes place but does not produce any spectral modification.
At lower densities two further resonances can occur that depend on $(\theta_{13}, \delta m^2_{13})$ and $(\theta_{12}, \delta m^2_{12})$, usually termed as the high resonance and the low resonances. The sign of the neutrino mass-squared differences determines if neutrinos or anti-neutrinos undergo a resonant conversion. The sign of $ \delta m^2_{12}$ produces a low resonance in the neutrino sector. The one of $ \delta m^2_{13}$ keeps unknown (the hierarchy problem). 
The adiabaticity of the evolution at the resonances depends also on the neutrino energy and on the gradient of the matter density which is fulfilled for typical power laws that accord with simulations  \cite{Dighe:1999bi}.  

Recent calculations have shown the emergence of new phenomena due to the neutrino-neutrino interaction, the presence of shock waves and of turbulence (see \cite{Duan:2010bg,Duan:2009cd} for a review).  Steep changes of the stellar density profile due to shock waves induce multiple MSW resonances and interference phenomena among the matter eigenstates. As a consequence the neutrino evolution can become completely non-adiabatic
when the shock passes through the MSW region.

As for the neutrino self-interaction it can produce  collective stable and unstable modes of the (anti-)neutrino gas and a swapping of the neutrino fluxes  with spectral changes. 
Various models have been studied to investigate the impact of the self-interactions on the neutrino spectra and the occurrence of collective instabilities that trigger flavour modifications in the star.
The first model, so called "bulb" model, was assuming that the spherical and azymuthal symmetries for neutrino propagation from the neutrino sphere, homogeneity and stationarity\footnote{The neutrino sphere is the flavour and energy dependent location deep inside the star from which neutrinos start free-streaming.}.
Within this model three flavor conversion regimes are present and well understood (the synchronisation, the bipolar oscillations and the spectral split). 
For example the spectral split phenomenon is an MSW-effect in a comoving frame \cite{Raffelt:2007xt}, or analogous to a magnetic-resonance phenomenon \cite{Galais:2011gh}
(see \cite{Duan:2010bg} and references therein). 

The interplay between matter and neutrino self-interaction effects needs to be accurately considered. In fact matter can decohere the collective neutrino modes since neutrinos with different emission angles (in the so-called "multiangle simulations") at the neutrinosphere have different flavor histories \cite{EstebanPretel:2008ni}.  It appears as for now that simulations based on realistic density profiles from supernova one-dimensional simulations suppress neutrino self-interaction effects. However this is no longer true if non-stationarity and inhomogeneity is considered : small scale seed perturbations can create large scale instabilities \cite{Capozzi:2016oyk}.
One should keep in mind that the solution of the full dynamical problem should involve the seven dimensions $(\vec{x}, \vec{p}, t)$.   To make the problem numerically computable the models involve various approximations. These are usually non-stationarity, homogeneity, the spherical and/or azymuthal symmetries. However, it has been shown that even if initial conditions have some
symmetry, the solutions of the evolution equations does not necessarily retain it \cite{Raffelt:2013rqa}. 
To avoid the demanding solution of the equations, often the instabilities are     
determined by employing a linearised stability analysis \cite{Banerjee:2011fj,Vaananen:2013qja} (see e.g. \cite{Chakraborty:2015tfa}). Such analysis are useful to identify the location of the instability, while they do not inform of the spectral modifications.

The neutrino spectral swapping turned out to be significant in the context of the "bulb" model while they could well reveal minor modifications in simulations including non-stationarity, inhomogeneities and a realistic description of the neutrino sphere (see e.g.  \cite{Duan:2010bg}). For the latter, Ref.\cite{Sawyer:2015dsa} has in fact shown that fast conversions can occur very close to the neutrino sphere, even if mixings are not taken into account. 
Many general features are established, but important questions remain in particular on the conditions for the occurrence of flavor modifications and its impact on the neutrino spectra.

Another open question is the role of corrections beyond the usual mean-field in  the transition region. This is between the dense region within the neutrinosphere which is Boltzmann treated, to the diluted one outside the neutrinosphere where collective flavor conversion occurs. So far, this transition has been treated as a sharp boundary where the neutrino fluxes and spectra obtained in supernova simulations is used as initial conditions in flavour studies. 
Extended descriptions describing neutrino evolution in a dense medium have recently been derived using a coherent-state path integral \cite{Volpe:2013uxl}, the Born-Bogoliubov-Green-Kirkwood-Yvon hierarchy \cite{Volpe:2013uxl}, or the two-particle-irreducible effective action formalism \cite{Vlasenko:2013fja} (see also \cite{Sigl:1992fn,McKellar:1992ja}). 
Besides collisions, two kinds of corrections in an extended mean-field description are identified : spin or helicity coherence \cite{Vlasenko:2013fja}  and neutrino-antineutrino pairing correlations \cite{Volpe:2013uxl}. Such corrections are expected to be tiny, but the non-linearity of the equations could introduce significant changes of neutrino evolution in particular in the transition region. Numerical calculations are needed to investigate the role of spin coherence or neutrino-antineutrino pairing correlations or of collisions. A first calculation in a simplified model shows that helicity coherence might have an impact \cite{Vlasenko:2014bva}. 

Neutrino flavor conversion also occurs in accretion disks around black holes \cite{Malkus:2012ts} and binary compact objects such as black hole-neutrons star and neutron star-neutron star mergers \cite{Zhu:2016mwa,Frensel:2016fge}. In particular, flavour modification can be triggered by a cancellation of the neutrino matter and self-interaction contributions in these scenarios.  This produces a resonant phenomenon called the neutrino-matter resonance \cite{Malkus:2012ts}. 

Another interesting theoretical development is the establishment of connections between neutrino flavor conversion in massive stars and the dynamics, or behaviour, of many-body systems in other domains. Using algebraic methods, Ref.\cite{Pehlivan:2011hp} has shown that the neutrino-neutrino interaction Hamiltonian can be rewritten as a (reduced) Bardeen-Cooper-Schrieffer (BCS) Hamiltonian for superconductivty \cite{Bardeen:1957mv}. As mentioned above, Ref.\cite{Volpe:2013uxl} has included neutrino-antineutrino correlations of the pairing type which are formally analogous to the BCS correlations. The linearisation of the corresponding neutrino evolution equations  has highlighted the formal link between stable and unstable collective neutrino modes and those in atomic nuclei and metallic clusters \cite{Vaananen:2013qja}.

\subsection{Supernova neutrino observations}
\noindent
The observation of the neutrino luminosity curve from a future (extra)galactic explosion would closely follow the different phases of the explosion furnishing a crucial test of
supernova simulation predictions, and  information on the star and unknown neutrino properties. In particular, the occurrence of the MSW effect in the outer layer of the star
and collective effects depends on the value of the third neutrino mixing angle and the neutrino mass ordering. The precise measurement of the last mixing angle \cite{Abe:2011fz,An:2012eh,Ahn:2012nd} reduces the number of unknowns. Still, the neutrino signal from a future supernova explosion could tell us about the mass ordering, either from  the early time signal in ICECUBE \cite{Serpico:2011ir},  or by measuring the positron time and energy signal, in Cherenkov or scintillator detectors, associated with the passage of the shock wave in the MSW region \cite{Gava:2009pj}. Several other properties can impact the neutrino fluxes such as the neutrino magnetic moment \cite{deGouvea:2012hg}, non-standard interactions, sterile neutrinos. CP violation effects from the Dirac phase exist but appear to be small \cite{Balantekin:2007es,Gava:2008rp,Kneller:2009vd,Pehlivan:2014zua}. In spite of the range of predictions, the combination of future observations from detection channels with different flavor sensitivities, energy threshold and time measurements can pin down degenerate solutions and bring key information to this domain (see e.g. \cite{Vaananen:2011bf}). 

The SuperNova Early Warning System (SNEWS) and numerous other neutrino detectors around the world can serve as supernova neutrino observatories if a supernova blows up in the Milky Way, or outside our galaxy. Large scale detectors based on different technologies \cite{Scholberg:2012id} including liquid argon, water Cherenkov and scintillator are being considered. Upcoming observatories are the large scale scintillator detector JUNO \cite{An:2015jdp} and hopefully the water Cherenkov Hyper-Kamiokande \cite{Yano:2016rkf}.
These have the potential to detect neutrinos from a galactic and an extragalactic explosion as well as to discover the diffuse supernova neutrino background produced from supernova explosions up to cosmological redshift of 2. The latter could be observed by EGADS, i.e. the Super-Kamiokande detector with the addition of Gadolinium  \cite{Xu:2016cfv} (for a review see \cite{Beacom:2010kk,Lunardini:2010ab}).

\section{Ultra-high energy neutrinos}
The main mission of  high-energy neutrino telescopes is to search for galactic and extra-galactic sources of high-energy neutrinos to elucidate the source of cosmic rays and the astrophysical mechanisms that produce them. These telescopes also investigate neutrino oscillations, dark matter and supernova neutrinos (for IceCube). The 37 events collected in ICECUBE, with deposited energies ranging from 30 to 2 PeV,  is consistent with the discovery of high energy astrophysical neutrinos at 5.7 $ \sigma$ \cite{Aartsen:2014gkd}.  The 2 PeV event is the highest-energy neutrino ever observed.

High-energy neutrino telescopes are currently also providing data on neutrino oscillations measuring atmospheric neutrinos, commonly a background for astrophysical neutrino searches.   Using low energy samples, both ANTARES \cite{AdrianMartinez:2012ph} and IceCube/DeepCore \cite{Gross:2013iq} have measured the parameters $\theta_{23}$ and $ \Delta m^2_{23}$ in good agreement with existing data. ORCA \cite{Adrian-Martinez:2016fdl} and PINGU  \cite{TheIceCube-Gen2:2016cap}, IceCube extension in the 10 GeV energy range, could measure the mass hierarchy by exploiting the occurrence of the matter effect from neutrinos, both from the MSW and the parametric resonance occurring in the Earth \cite{Akhmedov:1998ui,Petcov:1998su}. Neutrino telescopes are also sensitive to other fundamental properties such as Lorentz and CPT violation  \cite{Abbasi:2010kx}, or sterile neutrinos. 
 
\begin{figure}
\begin{center}
\includegraphics[width=0.7\textwidth]{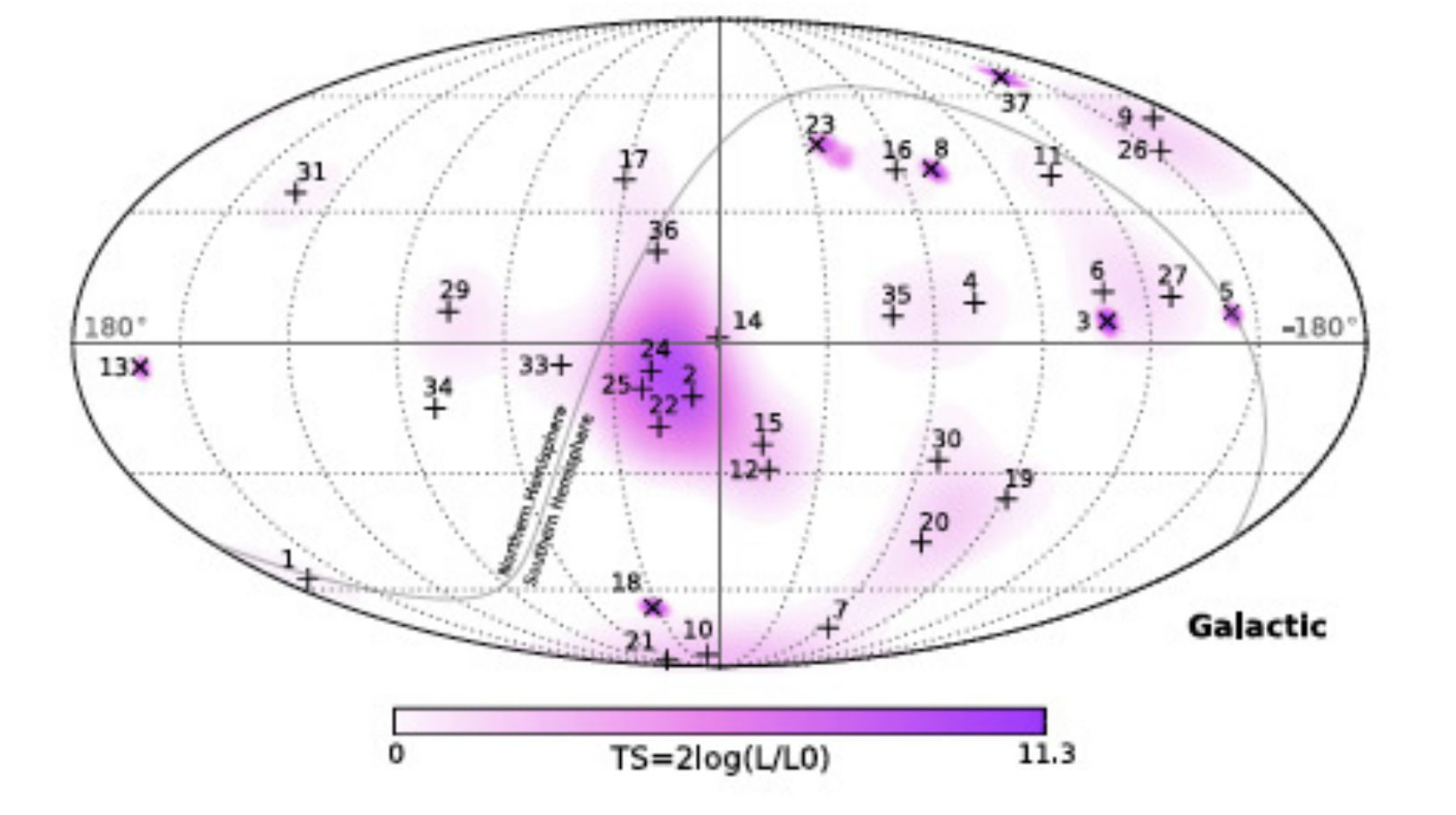}
\caption{}
\label{fig1}
\end{center}
\end{figure}

\end{document}